\documentclass[superscriptaddress,A4]{revtex4} 
\usepackage{psfig} 
\usepackage{graphicx} 
\usepackage[latin1]{inputenc} 
\usepackage{rotating} 
\newcommand{\EVRY}{D\'epartement de Physique et Mod\'elisation, 
Universit\'e d'Evry Val d'Essonne\\ 
Boulevard F. Mitterrand, 91025 Evry cedex} 
\newcommand{\LKB}{Laboratoire Kastler Brossel UMR 8552, EVRY, UPMC, CNRS, ENS, C 74, Universit\'e Pierre et Marie Curie\\ 
4 place Jussieu, 75252 Paris, France} 
\newcommand{\LPL}{Laboratoire de Physique des Lasers, UMR 7538 CNRS, Universit\'e Paris 13, 99 av. J.-B. Cl\'ement, 93430 Villetaneuse, France} 
\newcommand{\ILP}{The Institute of Laser Physics, Siberian Branch of the Russian Academy of Science, Pr. Lavrentyeva 13/3, 630090 Novosibirsk, Russia}
\begin{document} 
\title{HCOOH high resolution spectroscopy in the 9.18$\mu$m region}
\author{Franck Bielsa} 
\affiliation{\LKB} 
\affiliation{\EVRY}
\author{Khelifa Djerroud} 
\affiliation{\LPL}
\author{Andrei Goncharov}
\affiliation{\LPL}
\affiliation{\ILP}  
\author{Albane Douillet} 
\affiliation{\LKB} 
\affiliation{\EVRY}
\author{Tristan Valenzuela} 
\affiliation{\LKB} 
\affiliation{\EVRY}
\author{Christophe Daussy} 
\affiliation{\LPL}
\author{Laurent Hilico} 
\affiliation{\LKB} 
\affiliation{\EVRY}
\author{Anne Amy-Klein} 
\affiliation{\LPL}

\date{today} 
\begin{abstract} 
We report on higly accurate absolute frequency measurement against a femtosecond frequency comb of 6 saturated absorption lines of formic acid (HCOOH) with an accuracy of 1~kHz. We also report the frequency measurement of 17 other lines with an accuracy of 2~kHz. Those lines are in quasi coincidence with the 9R(36) to 9R(42) CO$_2$ laser emission lines and are probed either by a CO$_2$ or a widely tunable quantum cascade laser phase locked to a master CO$_2$ laser. 
The relative stability of two HCOOH stabilized lasers is characterized by a relative Allan deviation of 4.5~10$^{-12}$~$\tau^{-1/2}$. They give suitable frequency references for H$_2^+$ Doppler free two-photon spectroscopy.
\end{abstract}  
\maketitle


\section{introduction} 
Electron to proton mass ratio is a fundamental constant that is determined with a relative accuracy of 4.6~10$^{-10}$~\cite{codata} from separate electron and proton mass measurements~\cite{beier,farnham}. 
Molecular hydrogen ion (H$_2^+$ or HD$^+$) vibrational spectroscopy has recently been proposed as a new tool for very accurate direct optical
determination of the electron to proton mass ratio with an improved accuracy~\cite{hilico,karr,roth-schiller}. 
H$_2^+$ high resolution vibrational spectroscopy is
feasible using Doppler free two-photon transitions in the 9.1-9.2~$\mu$m range (1087-1099~cm$^{-1}$), corresponding to the 9R(34)-9R(52) CO$_2$ laser emission lines.
There is no close coïncidence between CO$_2$ laser lines and H$_2^+$ two-photon lines. However,  
we have shown that the recently  developped quantum cascade laser sources (QCL) are suitable for H$_2^+$ spectroscopy.
The relative accuracy level of interest on the vibrational frequency determination for mass ratio metrology purposes is 2.10$^{-10}$, 
that corresponds to 2.~10$^{-7}$~cm$^{-1}$ (6~kHz), hence the need for highly accurate molecular frequency references in the 9.1-9.2~$\mu$m range.

\begin{figure}[t] 
\center 
\includegraphics[width=8cm, angle=-90]{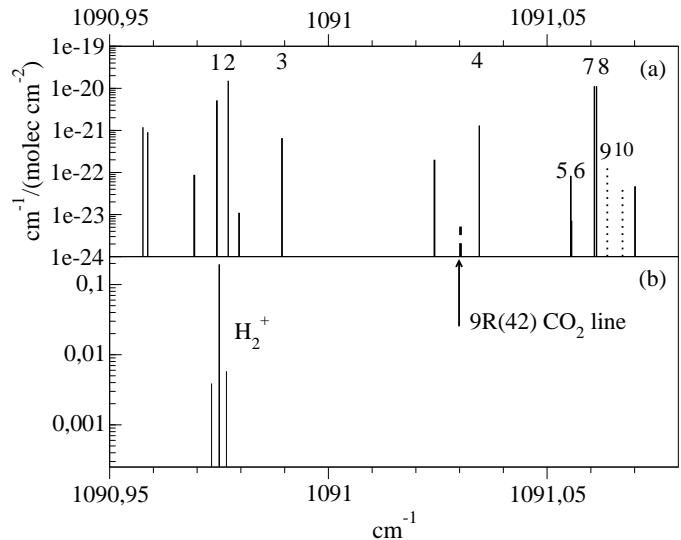} 
\caption{\label{spectrum} (a) Full lines : HCOOH lines of intensity larger than 10~$^{-23}$~cm$^{-1}$/(molec cm$^{-2}$) from HITRAN database~\cite{HITRAN}. Dotted lines : additional lines predicted by A.~Perrin~\cite{perrin}. The labels indicate the presently measured lines of Table~\ref{Measures}. When operated on the 9R(42) CO$_2$ line, the CO$_2$/HCOOH reference laser is stabilized on line 4.(b) (v=0,L=2)$\rightarrow$(v=1,L=2) H$_2^+$ two-photon transition spectrum in arbitrary units.} 
\end{figure}

Recently, the 9~$\mu$m band $^{12}$C$^{16}$O$_2$ absolute frequencies have been measured with an accuracy better than 1~kHz up to the 9R(36) line~\cite{amy}
using a CO$_2$ saturated absorption stabilized CO$_2$ laser.  Because the 9~$\mu$m band is a hot band and because the molecular population dramatically decrease with the angular quantum number, CO$_2$ saturated absorption signal to noise ratio becomes too low for laser stabilisation purposes beyond the 9R(36) line.
Among the large variety of molecules (SF$_6$, OsO$_4$, NH$_3$, CO$_2$, HCOOH) which are well known to provide frequency references for the CO$_2$ lasers,
formic acid $\nu_6$ band is the only one that presents intense lines in the 9.1-9.2~$\mu$m range~\cite{HITRAN}.
As an example, Fig.~\ref{spectrum} shows the (v=0,L=2)$\rightarrow$(v=1,L=2) H$_2^+$ two-photon excitation spectrum and formic acid lines close to the 9R(42) CO$_2$ emission line.
Through the paper, CO$_2$ and HCOOH denote the standard isotopes. At 300~K, HCOOH is in the {\it trans} conformation, the {\it cis} one representing only 0.1\% of the molecules~\cite{Vander}.

Line frequency measurements in the HCOOH $\nu_6$ band have been performed using Fourier transform, diode laser, CO$_2$ laser, and laser-radiofrequency double resonance and microwave spectroscopy~\cite{Willemot,Landsberg,Man,Bumgarner,Tan,baskakov}.
Among them, the most accurate ones were obtained by CO$_2$ laser spectroscopy with accuracies in the 50-250~kHz range~\cite{Landsberg} for a small number of lines. Recently, absolute line intensities of the $\nu_6$ and $\nu_8$ band have been determined accurately~\cite{Vander}. When compared to theoretical models taking into account Coriolis coupling between $\nu_6$ and $\nu_8$ bands, those measurements lead to the determination of effective hamiltonian parameters and the prediction of HCOOH lines with an uncertainty of 10$^{-6}$~cm$^{-1}$ (30~kHz)~\cite{HITRAN}. Those predictions have recently been improved, the line strengh being corrected by a factor 2 and the line frequency given with the same uncertainty.

The aim of this work is the measurement at the kHz accuracy level of molecular reference frequencies in quasi coincidence with the 9R(36) to 9R(42) CO$_2$ laser lines. We first describe the CO$_2$/HCOOH reference laser and the widely tunable QCL we have implemented. We then describe the frequency measurements and give the absolute frequencies of 23 HCOOH lines. Finally, we analyse the relative stability of two independant lasers locked to two different HCOOH lines.

\section{The laser sources}
The experimental setup consists in a HCOOH stabilised CO$_2$ reference laser (see Fig.~\ref{fig-co2}) and in a tunable QCL (see Fig.~\ref{fig-qcl}).
\subsection{The CO$_2$/HCOOH reference laser} 
The CO$_2$ laser is a sealed-off low pressure (13~Torr) dual-discharge-arm 1m-long laser emitting in a single longitudinal mode. The laser cavity is made of a 90-95\% efficiency diffraction grating in the Littrow configuration and of a high reflectivity mirror at 9.2$\mu$m. The main output of the laser is the zero$^{th}$ order of the grating. Laser oscillation is obtained up to the 9R(48) emission line of CO$_2$ at 9.142$\mu$m with more than 1W of optical power. 
The CO$_2$ laser frequency is stabilised onto a formic acid saturated absorption signal detected in transmission of a 1m long optical cavity of finesse 100 and beam waist 3.8~mm containing a low pressure HCOOH gaz (0.5-2 mTorr, 66-266 Pa). Both the laser and the cavity mirrors are glued on piezoelectric transducers for sweeping and modulation purposes. The cavity resonance frequency is sligthly modulated (frequency: f$_1$=33~kHz, depth: $\sim$1~kHz) and is locked to the laser frequency using first harmonic phase sensitive detection at f$_1$ of the transmitted intensity. The intracavity optical power is set at 400$\mu$W.
The laser frequency is modulated (frequency f$_2$=5.67~kHz, depth 100-200~kHz). A third harmonic phase-sensitive detection at 3f$_2$ gives the saturated absorption signal (see Fig. \ref{line}) with a high signal to noise ratio of more than 1000 at 1s integration time. This signal is used to stabilise the CO$_2$ laser with a loop bandwidth of 100Hz. In the case of the 9R(42) CO$_2$ line, there is no coincidence with intense HCOOH lines, so the CO$_2$ laser frequency is up-shifted by 128~MHz using an acousto-optic modulator to reach a HCOOH line.
\subsection{The QCL system} 
The QCL is a singlemode cw distributed feedback laser~\cite{model} operated in a liquid nitrogen cryostat. The threshold current is 400~mA at 77~K, the maximum current is about 1~A with a 9~V polarisation voltage. It can deliver up to 160mW optical power and is tunable from 9.166~$\mu$m to 9.240~$\mu$m. The temperature and current tunabilities are respectively 3~GHz/K and 150~MHz/mA. Because the QCL's emission spectrum is several MHz wide and exhibits a jitter over tens of MHz, the free running QCL is not suitable for high resolution infrared spectroscopy.
The spectral features of singlemode CO$_2$ lasers are much better with a linewidth in the kHz range~\cite{frech}. 
We have shown that 
they can be written out on the QCL spectrum with a 300-1500~MHz tunable frequency offset  using a fast phase-lock loop~\cite{bielsa} with a bandwidth of 6~MHz. The tunability range has been extended to 2~GHz by applying a 3-5~V reverse bias to the HgCdZnTe room temperature detector that monitors the beat note between the CO$_2$ laser and the QCL, resulting in a better detectivity.
The 6~MHz loop bandwidth implies that the CO$_2$ laser frequency modulation is also written out on the QCL frequency with the same depth. 
For HCOOH spectroscopy, the QCL beam is injected into a second Fabry-Perot cavity similar to the first one (1.5~m long, finesse 100, beam waist 4.2~mm, intra-cavity optical power 400~$\mu$W, HCOOH pressure 0.5-2~mTorr). The cavity resonance frequency is modulated (frequency f'$_1$=50~kHz, depth $\approx$1~kHz) and is locked to the QCL emission frequency. A third harmonic phase-sensitive detection at 3f$_2$ gives the HCOOH saturated absorption signal depicted in Fig.~\ref{line}.

\begin{figure}[h] 
\center 
\includegraphics[width=8cm]{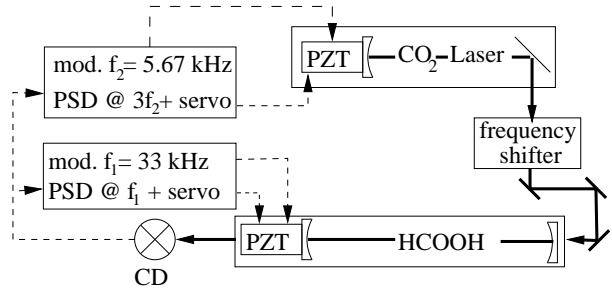} 
\caption{\label{fig-co2}CO$_2$/HCOOH reference laser. The CO$_2$ laser is stabilised on a HCOOH saturated absorption line. The frequency shifter is an acousto-optic modulator only used when the CO$_2$ laser is operated on the 9R(42) CO$_2$ line to up-shift the laser frequency by 128 MHz. CD: Liquid nitrogen cooled HgCdTe detector. PZT: piezo transducer. PSD : phase sensitive detection.} 
\end{figure} 
\begin{figure}[h] 
\center 
\includegraphics[width=8cm]{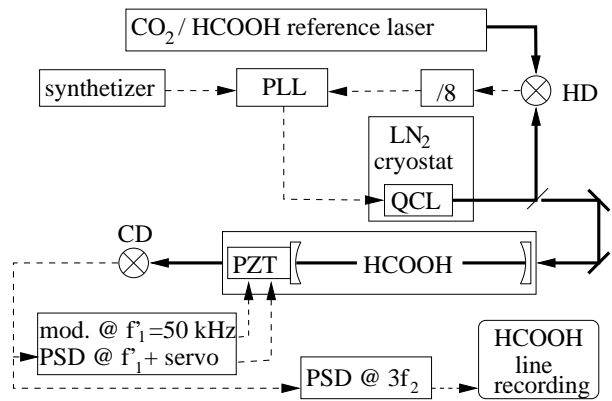} 
\caption{\label{fig-qcl}QCL spectrometer. The QCL and CO$_2$/HCOOH reference laser beams are beated on a fast room temperature HgCdZnTe detector (HD). The beatnote frequency is divided by 8 and phase compared to a RF synthesizer of frequency $\delta$. The QCL is phase-locked to the reference laser with a tunable frequency offset $8\delta$. Because the frequency modulation of the reference laser is transferred to the QCL by the fast phase-lock loop, the HCOOH saturated absorption line is phase sensitive detected at the third harmonic f$_2$, see Fig.~\ref{fig-co2}. PZT: piezo transducer, PLL : phase-lock loop.} 
\end{figure}
\section{Results and discussion}
We have performed absolute (A) and relative (R) frequency measurements of HCOOH saturated absorption lines. The results are given in Table~\ref{Measures} and are respectively labelled by A or R in the second column. 
\subsection{\label{abs}Absolute frequency mesurement} 
For the absolute measurements, the CO$_2$/HCOOH reference laser (Fig.~\ref{fig-co2}) has been transported to LPL laboratory where a femtosecond frequency comb infrared frequency measurement facility is implemented~\cite{amy-2005}. We have measured the absolute frequency of 6 HCOOH lines in quasi coincidence with 9R(36) to 9R(42) CO$_2$ laser lines for which no accurate molecular reference were known. 

The HCOOH/CO$_2$ reference laser frequency is measured by comparison with a very high-harmonic of the repetition 
rate of a femto-second (fs) laser~\cite{amy-2005}. In the frequency domain, the fs laser provides a 
comb of modes separated by the repetition rate~\cite{cundiff}. A second comb is 
produced by a sum-frequency generation (SFG) of the fs laser comb and the CO$_2$ 
laser in a nonlinear crystal. This results in a beat between the low frequency part of the SFG comb and the 
high frequency part of the initial comb. The infrared frequency is thus compared 
to the difference between two modes of the comb. The beatnote is finally used to 
phase-lock the repetition rate to the CO$_2$ laser frequency. This scheme is independent 
of the comb offset and does not require any broadening of the comb.
To complete the frequency measurement procedure, the repetition rate (about 1 GHz) is detected 
with a fast photodiode and counted against a local oscillator at 1 GHz. This local oscillator 
is phase-locked to a reference signal transmitted via a 43-km long optical fiber from the LNE-SYRTE 
laboratory, located in Paris~\cite{daussy}. This laboratory has developed a high stability oscillator, 
which is based on a combination of a cryogenic Sapphire oscillator (CSO), an H-Maser and a set of 
low noise microwave synthesizers~\cite{chambon}. Its frequency is steered by the H-Maser in the 
long-term, and monitored by the Cs atomic fountain for accuracy~\cite{vian}. This signal shows 
a frequency stability slightly below 
$10^{-14}\ \tau^{-1}$ 
in the range 1-10~s, and 10$^{-15}$ 
from 10 to 105 s. The transfer through the optical link degrades this stability by less than one 
order of magnitude, while the phase noise introduced by the link can be efficiently suppressed 
with an active correction.
Finally the resolution of the whole measurement chain is better than 0.3 Hz (10$^{-14}$ in 
relative value) and will definitely not be a limitation for the present measurements.

The day to day repetability of the measurements is 100~Hz for controlled experimental 
conditions. No significant frequency shifts are observed by varying the intracavity 
optical power between 400$\mu$W and 2mW, varying the laser modulation depth by more 
than a factor eight and introducing significant but small error signal offsets. 
We have observed a positive pressure shift smaller than 2~kHz by varying the HCOOH 
pressure up to 4~mTorr. At the 0.5~mTorr level, formic acid pressure cannot be 
precisely controlled in our setup, probably because of adsorption effects on the 
vacuum chamber walls. We therefore give a conservative uncertainty of 1~kHz or 3.3~10$^{-8}$~cm$^{-1}$ on 
the absolute frequency measurements.

\subsection{HCOOH spectroscopy} 
\begin{figure}[h]
\center 
\includegraphics[width=8cm, angle=0]{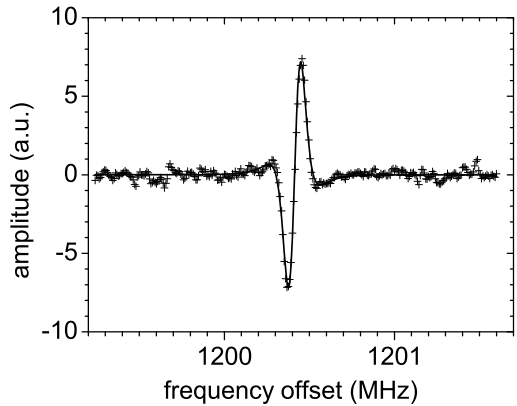} 
\caption{\label{line} Third harmonic phase sensitive saturated absorption signal of HCOOH line n°~3 
of Table~\ref{Measures} versus the frequency offset 8~$\delta$ between the reference laser and 
the phase locked QCL (see Fig.~\ref{fig-qcl}). The integration time is 1~s per point. Full line: adjustement 
by a third lorentzian derivative. The full width at half maximum is 220~kHz.} 
\end{figure}

We have measured the HCOOH lines that are red or blue detuned by 300-1600~MHz with 
respect to the closest CO$_2$/HCOOH  reference laser line using the widely tunable 
phase-locked QCL. The QCL frequency is swept by 1.6~kHz increments and the saturated 
absorption profile is recorded and fitted by the third derivative of a lorentzian 
shape to determine the line center and width (see Fig~\ref{line}). We are easily 
able to detect lines with a signal to noise ratio between 3 and 1000 in a 1s integration 
time, the former corresponding to normalized intensities down to 
1.6~10$^{-22}$~cm$^{-1}$/(molec cm$^{-2}$), (taking into account the recent corrections of the line intensities by a factor of two~\cite{Vander}). With optimized intracavity power and 
gaz pressure, the sensitivity of the spectrometer is low enough to detect 
the CO$_2$ 9R(42) saturated absorption line that 
have a 5.~10$^{-24}$~cm$^{-1}$/(molec cm$^{-2}$) normalized intensity.
Nevertheless, absolute HCOOH line intensities cannot be determined precisely 
from the line profile because of the large uncertainty on the actual pressure in the cavity.

The HCOOH fitted linewidths are between 160 and 220~kHz and are dominated by modulation broadening. 
Indeed, the transit time broadening is 26~kHz, the pressure broadening is 0.32~cm$^{-1}$/atm~\cite{Vander} i.e. 
12.8~kHz/mTorr and the natural width and power broadening are negligible.

Thanks to the high signal to noise ratio of the lines, the fitted line centers are determined 
with uncertainties better than 1~kHz.
Nevertheless, taking into account the independant uncertainties 
on the reference line frequency (1~kHz), the line position (1~kHz) 
and the HCOOH pressure determination (1~kHz),
we give a conservative uncertainty of 2~kHz or 6.6~10$^{-8}$~cm$^{-1}$ on the 
HCCOH line frequency determined with the QCL spectrometer. 
\subsection{Discussion}
In Table~\ref{Measures}, the measured frequencies are compared to the HITRAN database predicted values~\cite{HITRAN}, that are given with a 10$^{-6}$~cm$^{-1}$ resolution. 
Our measurements agree with the HITRAN predictions at the 10$^{-4}$~cm$^{-1}$ level for most of the lines excepted one for which a 8.~10$^{-4}$~cm$^{-1}$ (24~MHz) offset is observed.
The agreement with recently predicted lines by A. Perrin~\cite{perrin} is better at the 2.~10$^{-5}$~cm$^{-1}$ level excepted for two lines.
In the last column of the Table, the $\ast$ labels predicted doublets. 

A few doublets predicted by A. Perrin~\cite{perrin} at 1089.040400, 1089.998920, 1090.055270 and 1091.066960~cm$^{-1}$ have not been observed in a frequency interval of $\pm$~8~MHz (2.6~10$^{-4}$~cm$^{-1}$) around the predicted values, with a sensitivity of 10$^{-22}$~cm$^{-1}$/(molec cm$^{-2}$).

In Table~\ref{comparison}, we compare the present frequency measurements with previously published ones. There is a good agreement with the measurements obtained by CO$_2$ laser saturation spectroscopy and reported in~\cite{Landsberg}, the accuracy being improved by at least one order of magnitude.  
\begin{table} 
\small
\begin{tabular}{|c|c|c|c|l|r|r|ccccccc|l|c|} 
\hline
CO$_2$&&f&$\delta f$&\multicolumn{1}{c|}{$\sigma_m$}&\multicolumn{1}{c|}{$\delta_H$}&\multicolumn{1}{c|}{$\delta_P$}&&&&&&&&&\\ 
line&&(kHz)&(kHz)&\multicolumn{1}{c|}{(cm$^{-1}$)}&\multicolumn{1}{c|}{(cm$^{-1}$)}&\multicolumn{1}{c|}{(cm$^{-1}$)}&\multicolumn{7}{c|}{\begin{tabular}{*{7}{p{0.4cm}}}&J'&K$_a$'&K$_c$'&J''&K$_a$''&K$_c$''\end{tabular}}&&\\ 
\hline 
9R(36)&A&32 615 874 282.1&1.0&1087.948 459 40&&-0.000601&\multicolumn{7}{c|}{\begin{tabular}{*{7}{p{0.4cm}}}V6&26&21&5&27&21&6\\
                                                                                                          V6&26&21&6&27&21&7
										\end{tabular}}&$\ast$&\\		  
\hline 
9R(38)&R&32 645 920 103.8&2.0&1088.950 680 13&-0.000030
&0.000040&\multicolumn{7}{c|}{\begin{tabular}{*{7}{p{0.4cm}}}V6&17&3&15&17&4&14\end{tabular}}&&\\ 
    &R&32 646 482 873.7&2.0&1088.969 452 12&0.000022
    &0.000072&\multicolumn{7}{c|}{\begin{tabular}{*{7}{p{0.4cm}}}V6&19&3&16&20&3&17\end{tabular}}&&\\ 
    &R&32 646 774 027.4&2.0&1088.979 163 96&-0.000216
    &-0.000096&\multicolumn{7}{c|}{\begin{tabular}{*{7}{p{0.4cm}}}V6&21&11&10&22&11&11\\
                                                                V6&21&11&11&22&11&12
                                                                                            \end{tabular}}&$\ast$&\\ 
    &A&32 647 442 530.0&1.0&1089.001 462 80&-0.000247&-0.000117&\multicolumn{7}{c|}{\begin{tabular}{*{7}{p{0.4cm}}}V6&30&1&29&30&2&28\end{tabular}}&&\\ 
    &A&32 647 451 806.2&1.0&1089.001 772 23&&-0.000408&\multicolumn{7}{c|}{\begin{tabular}{*{7}{p{0.4cm}}}V6&52&6&46&53&5&49\end{tabular}}&&\\ 
    &R&32 648 078 880.7&2.0&1089.022 689 18&-0.000031&-0.000051&\multicolumn{7}{c|}{\begin{tabular}{*{7}{p{0.4cm}}}V6&16&3&14&16&4&13\end{tabular}}&&\\ 
    &R&32 648 797 827.9&2.0&1089.046 670 68&0.000101&-0.000019&\multicolumn{7}{c|}{\begin{tabular}{*{7}{p{0.4cm}}}V6&20&0&20&21&0&21\end{tabular}}&&\\ 
\hline 
9R(40)&R&32 677 093 109.1&2.0&1089.990 499 66&0.00007&-0.000010&\multicolumn{7}{c|}{\begin{tabular}{*{7}{p{0.4cm}}}V6&29&1&28&29&2&27\end{tabular}}&&\\  
    &R&32 677 848 146.1&2.0&1090.015 684 99&0.000105&-0.000255&\multicolumn{7}{c|}{\begin{tabular}{*{7}{p{0.4cm}}}V6&19&1&19&20&0&20\end{tabular}}&&\\ 
    &A&32 678 207 409.6&1.0&1090.027 668 73&-0.000021
    &-0.000021&\multicolumn{7}{c|}{\begin{tabular}{*{7}{p{0.4cm}}}V6&18&3&16&19&3&17\end{tabular}}&&\\ 
    &A&32 678 247 320.7&1.0&1090.029 000 02&&-0.000070&\multicolumn{7}{c|}{\begin{tabular}{*{7}{p{0.4cm}}}V6&22&0&22&22&2&21\end{tabular}}&&\\  
    &R&32 679 212 591.0&2.0&1090.061 197 97&0.000018&0.000088&\multicolumn{7}{c|}{\begin{tabular}{*{7}{p{0.4cm}}}V6&4&2&3&5&3&2\end{tabular}}&&\\
    &R&32 679 681 163.9&2.0&1090.076 827 88&0.000018&-0.000012&\multicolumn{7}{c|}{\begin{tabular}{*{7}{p{0.4cm}}}V6&4&2&2&5&3&3\end{tabular}}&&\\ 
\hline 
9R(42)&R&32 706 583 453.5&2.0&1090.974 191 67&-0.000338&0.000082&\multicolumn{7}{c|}{\begin{tabular}{*{7}{p{0.4cm}}}V6&18&9&9&19&9&10\\
                                                                                                        V6&18&9&10&19&9&11\end{tabular}}&$\ast$&1\\ 
    &R&32 706 672 889.5&2.0&1090.977 174 93&0.000025&-0.000035&\multicolumn{7}{c|}{\begin{tabular}{*{7}{p{0.4cm}}}V6&17&2&16&18&2&17\end{tabular}}& &2 \\
    &R&32 707 063 566.1&2.0&1090.990 206 50&0.000797&0.000077&\multicolumn{7}{c|}{\begin{tabular}{*{7}{p{0.4cm}}}V6&20&15&5&21&15&6\\
                                                                                                        V6&20&15&6&21&15&7\end{tabular}}&$\ast$&3\\ 
    &A&32 708 391 980.5&1.0&1091.034 517 64&-0.000038&-0.000032&\multicolumn{7}{c|}{\begin{tabular}{*{7}{p{0.4cm}}}V6&21&2&20&21&3&19\end{tabular}}& &4 \\     
    &R&32 709 015 560.0&2.0&1091.055 318 01&-0.000042&-0.000112&\multicolumn{7}{c|}{\begin{tabular}{*{7}{p{0.4cm}}}V6&8&4&5&7&5&2\end{tabular}}& &5 \\ 
    &R&32 709 017 250.3&2.0&1091.055 374 39&-0.000036&-0.000116&\multicolumn{7}{c|}{\begin{tabular}{*{7}{p{0.4cm}}}V6&8&4&4&7&5&3\end{tabular}}& &6 \\ 
    &R&32 709 175 798.5&2.0&1091.060 662 99&-0.000147&-0.000023&\multicolumn{7}{c|}{\begin{tabular}{*{7}{p{0.4cm}}}V6&17&5&12&18&5&13\end{tabular}}& &7 \\ 
    &R&32 709 189 841.6&2.0&1091.061 131 42&-0.000149&0.000081&\multicolumn{7}{c|}{\begin{tabular}{*{7}{p{0.4cm}}}V6&17&5&13&18&5&14\end{tabular}}& &8 \\ 
    &R&32 709 274 924.5&2.0&1091.063 969 48&&-0.000091&\multicolumn{7}{c|}{\begin{tabular}{*{7}{p{0.4cm}}}V6&22&20&3&23&20&4\\
                                                                                                          V6&22&20&2&23&20&3\end{tabular}}& &9 \\ 
\hline 
\end{tabular} 
\caption{\label{Measures}HCOOH line frequencies. The first column gives the closest CO$_2$ laser line and the second one the measurement system (A for absolute frequency measurement and R for relative measurement by QCL spectroscopy). Columns 3 and 4 give the measured frequencies f (this work) and their uncertainties $\delta f$. Column 5 is the corresponding wavenumber $\sigma_m$. Columns 6 and 7 give the differences $\delta_H=\sigma_m-\sigma_H$ and $\delta_P=\sigma_m-\sigma_P$ between our measurements and the wavenumbers $\sigma_H$ and $\sigma_P$ predicted respectively by the HITRAN database~\cite{HITRAN} and A. Perrin~\cite{perrin}.
 The following columns give the usual line nomenclature. 
($\ast$) denotes predicted doublets.  
Figures in last column refers to Fig.~\ref{spectrum}.} 
\end{table} 
\begin{table}
\begin{tabular}{|c|c|c|c|ccccccc|c|}
\hline
&(a)&(b)&(a)-(b)&&&&&&&&\\
\hline
&kHz&MHz&kHz&\multicolumn{7}{c|}{\begin{tabular}{*{7}{p{0.4cm}}}&J'&K$_a$'&K$_c$'&J''&K$_a$''&K$_c$''\end{tabular}}&ref.\\
\hline
9R(38)&32 646 774 027.4 (2.0)&32 646 775.1 &-1072.6&\multicolumn{7}{c|}{\begin{tabular}{*{7}{p{0.4cm}}}V6&21&11&10&22&11&11\\
                                                                                                 V6&21&11&11&22&11&12\end{tabular}}&\cite{Tan}\\
      &32 647 442 530.0 (1.0)&32 647 442.52(0.1)&10.0&\multicolumn{7}{c|}{\begin{tabular}{*{7}{p{0.4cm}}}V6&30&1&29&30&2&28\end{tabular}}&\cite{Landsberg} *\\
      &32 647 451 806.2 (1.0)&32 647 451.80(0.1)&6.2&\multicolumn{7}{c|}{\begin{tabular}{*{7}{p{0.4cm}}}V6&52&6&46&53&5&49\end{tabular}}&\cite{Landsberg} *\\
\hline
9R(40)&32 678 207 409.6 (1.0)&32 678 207.36&49.6&\multicolumn{7}{c|}{\begin{tabular}{*{7}{p{0.4cm}}}V6&18&3&16&19&3&17\end{tabular}}&\cite{Landsberg}\\
\hline
\end{tabular}
\caption{\label{comparison}Comparison of the present frequency measurements (a) and already published measurements (b). 
The line assignements are those given in HITRAN or the recent reference~\cite{perrin}. $\ast$: the line assignement in ref.~\cite{Landsberg} is different.}
\end{table}
\subsection{Reference frequency stability} 
\begin{figure}[h] 
\center 
\includegraphics[width=8cm]{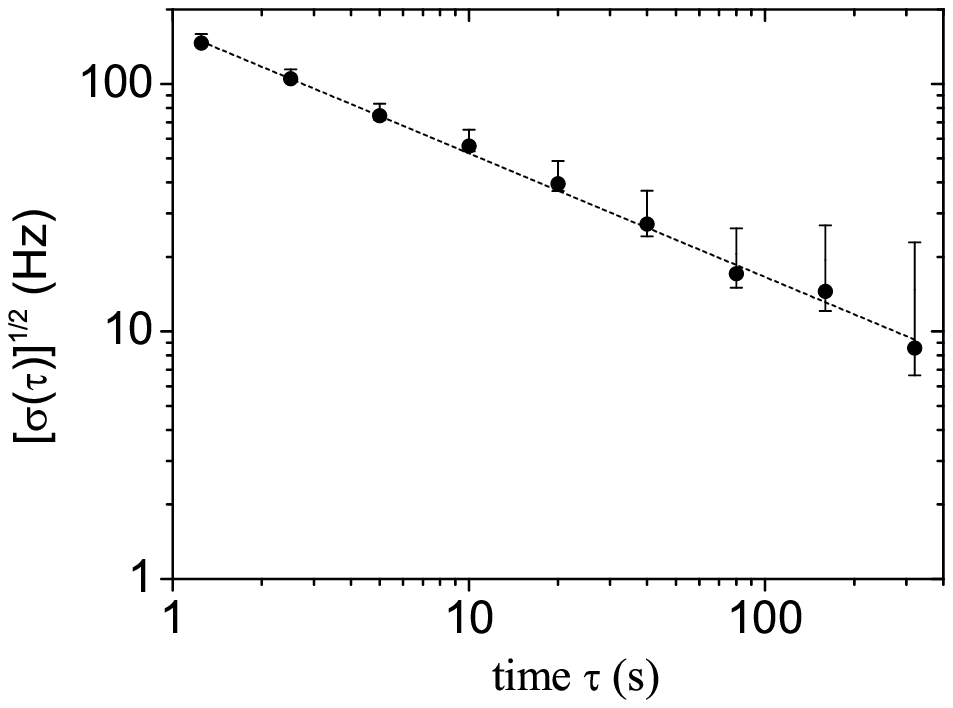} 
\caption{\label{ADEV} Square root of the Allan deviation of the two HCOOH stabilized laser beatnote as a function of integration time $\tau$, obtained from a record of 633 1s gate time measurements. The dashed curve is a fit giving a 146~Hz~$\tau^{-1/2}$ behaviour. Dead time between measurements is 0.25~s.} 
\end{figure}

The aim of this work is to realise a reliable and stable frequency reference to probe two-photon vibrational transitions in H$_2^+$ with a resolution of the order of 1~kHz.
We have measured the relative frequency stability of two independant systems. The first one is the CO$_2$/HCOOH reference laser stabilized on the HCOOH line n°4 of Table~\ref{Measures}. 
The second system involves the QCL and a master CO$_2$ laser similar to the first one. The QCL is phase locked to the CO$_2$ laser with a fixed frequency offset. The QCL beam probes HCOOH line n°7 of Table~\ref{Measures}. The saturated absorption signal is used to lock the master CO$_2$ laser frequency.
To characterize the relative stability of the two laser systems, the beatnote frequency is counted with 1~s gate time. The Allan deviation shown in Fig.~\ref{ADEV} exhibits a white frequency noise behaviour of 146~Hz~$\tau^{-1/2}$, that is 4.5~10$^{-12}$ in relative value. The relative stability reaches the 10~Hz level for 200~s integration time. Both lasers are locked using saturated absorption signal having similar signal to noise ratios so individual stabilities are expected to be comparable and of the order of 100~Hz~$\tau^{-1/2}$.
\section{Conclusion} 
We have set up a CO$_2$/HCOOH reference laser that can be operated from the 9R(36) to 9R(42) line and measured the absolute frequency of 6 saturated absorption lines with an accuracy of 1~kHz (3.3~10$^{-8}$cm$^{-1}$) improved by two orders of magnitude as compared to previously reported results or predictions. 
Using a quantum cascade laser phase (QCL) locked to a master CO$_2$ laser, we have determined the frequency of 17 additional formic acid lines with an accuracy of 2~kHz.
The frequency stability of both the reference laser and the phase locked QCL are characterized by a relative Allan deviation of 3.1~10$^{-12}$~$\tau^{-1/2}$. This performance demonstrates that  the CO$_2$/HCOOH phase locked QCL is a suitable tool for H$_2^+$ high resolution spectroscopy at the kHz level.

\begin{acknowledgments} 
We thank O. Acef, G. Santarelli and M. Lours (LNE-SYRTE). We also thank A. Perrin (LISA, Université de Créteil) for fruitfull discussions about HCOOH. 
Laboratoire Kastler Brossel is UMR 8552 du CNRS. This work was supported by an 
ACI jeune 03-2-379, BNM grants 02-3-008 and 04-03-009 and Evry University. 
\end{acknowledgments} 

\end{document}